\documentstyle[12pt]{article}
\begin{document}
\begin{center}
{\bf Specialists' Meeting on the Nucleon-Nucleus \\
Optical Model up to 200 MeV: \\
Conclusions and Recommendations}\\
\vspace{12pt}
D.\ G.\ Madland\\
\vspace{12pt}
{\it Theoretical Division, Los Alamos National Laboratory\\
Los Alamos, New Mexico 87545}
\end{center}
\noindent If I may be permitted to speak for the participants of this
meeting, we owe our very warm thanks and therefore a very strong round
of applause to the ORGANIZERS of the meeting: Olivier Bersillon, Jean
Paul Delaroche, Arjan J.\ Koning, and Pierre Nagel, to the HOSTS of the
meeting: the {\it Service de Physique et Techniques Nucl$\acute{e}$aires,
Centre d'$\acute{e}$tudes de Bruy$\grave{e}$res-le-Ch$\hat{a}$tel}, and to
our SECRETARY: Nadine Labbal. I believe that seldom, if ever, have we
simultaneously experienced such warmth and elegance at a scientific
gathering. We thank you and we salute you.
 
\vspace{12pt}
\noindent May I also say, for all of us here, that we remember with
respect our colleague here at B-III, Dr. Christian Lagrange, who
during his life did much to advance our knowledge and understanding
of the optical model.
 
\vspace{12pt}
\noindent Now what has our meeting been about?
Experimentalists and theorists who have been addressing the medium-energy
nucleon-nucleus optical model potential have been invited here to discuss
their work so that we can assess our present understanding and to determine
what next is to be done. All of this is driven by (a) the goals of fundamental
physics understanding and complete predictive power for the elastic
scattering observables and (b) the crucial importance of the optical
model to the nuclear reaction codes that treat simultaneously all
competing non-elastic channels.
In particular, accurate knowledge of the total
scattering flux and its partition into elastic and non-elastic components
is provided by a physically correct optical potential. This is the
very first step in obtaining a physically realistic evaluation of the complete
set of cross sections for all open channels at a given projectile energy.
The optical potential is thus a crucial ingredient in such evaluations
for the above reason and also because the scattering $S$ matrix that it
provides is itself an essential input to an array of nuclear reaction
codes that treat specific open channels. Therefore, the generation of
accurate medium-energy cross section and spectra libraries for applied
purposes depends critically upon the medium-energy nucleon-nucleus
optical potential.
 
\vspace{12pt}
\noindent So, what have we learned and concluded, and what recommendations
should be made? Considering experiment first, there exists a fairly extensive
medium-energy proton-nucleus scattering data base consisting of well-measured
differential elastic cross sections, asymmetries, and spin rotations.
However, there are relatively few proton total reaction cross section
measurements
at medium energies and many of these have large experimental uncertainties
[Lhenry]. Thus, a number of medium-energy proton total reaction cross section
measurements are needed at accuracies of (say) better than 5\%,
because this observable constrains the non-elastic scattering flux predicted
by the optical potential and, equally important, it discriminates between
two (or perhaps more) otherwise equivalent families of potentials
in a Dirac phenomenology approach [Madland].
The choice of targets and bombarding energies should, of course, be based
on the voids in the existing experimental data base and/or on complementarity
with existing measurements of the same (or similar) neutron elastic scattering
observables. Turning to medium-energy neutron-nucleus
elastic scattering observables, there exists a quite extensive and
well-measured total cross section data base [Finlay {\it et al.}], but very
few differential elastic cross sections and spin observables have been
measured. Accordingly, given the great difficulty of these latter
measurements,
a {\it few} well-chosen experiments should
nevertheless be performed in order to obtain complementarity with existing
measurements of the same (or similar) proton elastic scattering observables.
 
\vspace{12pt}
\noindent One recommendation, from the point of view of benchmarking
existing as well as new medium-energy nucleon-nucleus optical potentials,
is to consider two targets and three
nucleon bombarding energies: \{$^{40}$Ca, $^{208}$Pb; 100, 200, 300 MeV\}.
The neutron experiments would consist of measurements of six differential
elastic cross sections and six asymmetries, and these would have considerable
overlap with existing proton measurements of the same observables.
The proton experiments would consist of measurements of (a) six total
reaction cross sections and these would have unit overlap with existing
measurements of the neutron total cross sections (these are the integral
observables
that can be most accurately measured for the two probes).
[Note that the proton total reaction cross section for $^{40}$Ca has
already been reasonably well measured at 100 MeV], and
(b) three (p,n) quasi-elastic scattering cross sections to the isobaric analog
of the $^{208}$Pb target ground state.\footnote{Following the meeting
two of the invitees who were unable to attend [F.\
S.\ Dietrich and J.\ Rapaport] commented that if the (p,n) cross sections are
measured with and without polarized protons, ($A_{y}$), much more certain
information can be extracted on the real and imaginary parts
of the isospin-dependent (isovector) terms
of the interaction potential and their energy dependencies.}
The two targets both have sufficient numbers of nucleons to justify
the assumption of a
smooth absorptive potential to simulate the totality of the non-elastic
open channels. In addition, they are both doubly-closed major shell
nuclei, which allows unambiguous comparisons of the predictions
from Dirac and Schr\"odinger approaches.
Furthermore, $^{40}$Ca has $N = Z$ and zero isospin
whereas $^{208}$Pb has $N > Z$ and non-zero isospin, which allows separate
studies/tests with and without isospin dependence, at three different
bombarding energies
for the two nucleon probes. Finally, the two isotopically enriched targets
exist in amounts that are adequate for the suggested neutron experiments.
This set of measurements, together with the already completed complementary
measurements, would constitute a close approximation to a complete
and ideal set of measured scattering observables for optical-model
studies/tests of
medium-energy nucleon-nucleus elastic scattering as defined for this meeting.
 
\vspace{12pt}
\noindent A second recommendation, from the point of view of prioritizing
measurements for optical potentials important to medium-energy
applications, is to examine the High-Priority Nuclear Data Request
List for Intermediate Energies for the items indexed to the optical
model. Note that the List is periodically updated. It is located on
the web at http://www.nea.fr/html/trw/nucdat/iend/docs/doc.s13.21.html
[Koning].
 
\vspace{12pt}
\noindent Turning to theory and modeling,
a number of shared opinions on these topics surfaced during
this meeting. These include the views that (1) for medium-energy work it
is time to stop extrapolating the low-energy, widely-used and
respected, optical-model potentials to medium energies and to instead view
these potentials as low-energy boundaries in the development of
new medium-energy potentials, (2) conclusions from important microscopic
medium-energy optical-model studies
should strongly influence the development of phenomenological medium-energy
potentials, and (3) the medium-energy optical-model potential user community
is especially attracted to global phenomenological potentials because they are
so easily hardwired and used in an optical-model routine.
These views share the common denominator that we have not yet achieved
the goal of a satisfactory medium-energy optical potential (which is
why we are having this meeting in the first place).
 
\vspace{12pt}
\noindent Five general approaches to the medium-energy optical potential have
been discussed.
These are (a) Dirac and relativistic Schr\"odinger phenomenological potentials
[Madland, Ishibashi],
(b) dispersive potentials [Romain, Delaroche], (c) semi-microscopic
potentials [Bauge], (d) microscopic potentials [Elster, von Geramb],
and (e) coupled-channel potentials [Koning, Raynal].
Note that in some cases the medium-energy potential presented was
actually determined using two or more of these approaches simultaneously
[Bauge, for example]. While it is not the purpose here to summarize the many
excellent
technical presentations that we have heard, but to instead draw conclusions
and construct recommendations from them, it is nevertheless impossible not
to make
note of a few of the advances.
One of these is the extension of the Jeukene-Lejeune-Mahaux (JLM) folding
model upwards to 200 MeV [Bauge and Delaroche] by refitting the imaginary
part, introducing a phenomenological spin-orbit part, and employing
H-F-B densities calculated with the Gogny D1S force. A very careful and
systematic study of nucleon scattering by spherical nuclei then led to an
energy-dependent set of four potential-depth renormalization factors
\{$\lambda_{
pot}$\}. This work could be viewed as a first step in achieving a global
semi-microscopic medium-energy optical potential.
Another advance is the recognition of the discriminatory power of highly
accurate experimental neutron total cross sections in the various microscopic
approaches to the potential. For example, full-folding
calculations using the full Bonn $NN$ t-matrix and Dirac-Hartree densities
[Elster] as well as off-shell ``$t\rho$'' approximation calculations
[Picklesimer, Ernst]
are favored by the data, but calculations neglecting
the coupling of the struck target nucleon to the residual nucleus (a medium
effect) as well as those using a local, on-shell ``$t\rho$'' approximation
are rejected by the data.
Yet another advance is the extension of the dispersive approach to deformed
nuclei [Romain and Delaroche] providing a new way
to test and utilize the checks and balances existing between the bound and
continuum
(scattering) states and observables of a many-body deformed system.
Still another is the calculation [von Geramb] showing a strong sensitivity of
the $\pi\pi\;s$-wave scattering phase shift ($T=0$) to {\it small $\sim$ }2
MeV
changes in the pion mass, implying a resonance feature due to an in medium
effective pion mass. Since it is believed that correlated two pion exchange
is responsible for the attractive medium range $NN$ interaction this result
may constitute an important medium effect that should be quantified.
Finally, the manifestly interactive spherical optical model program ECISVIEW
[Koning] built upon ECIS-95 [Raynal] has the possibility to revolutionize
the traditional manner of obtaining a phenomenological potential
from experimental data, in both speed and (more importantly) completeness.
 
\vspace{12pt}
\noindent Based upon the technical presentations, subsequent discussions,
and roundtables at the end, some conclusions and recommendations on theory
and modeling follow:
\begin{enumerate}
\item On the question of Dirac {\it vs.} Schr\"odinger approaches to
medium-energy nucleon-nucleus scattering this still appears to be an
open question. The Schr\"odinger proponents point out that they have
carried out far more detailed investigations: medium effects, non-local
effects, off-shell effects, ..., and therefore should and do have (somewhat)
better agreement with experiment. Moreover, the Dirac agreement may
therefore be fortuitous because the Dirac community has not performed
such detailed investigations of these effects. The Dirac proponents
point out that the Dirac equation is the {\it correct} equation for
spin $\frac{1}{2}$ (point) fermions, that it has a {\it natural}
spin-orbit term, and that it has a {\it natural} Coulomb correction
term. Therefore, ``just give us time to address the detailed effects.''
Our recommendation is that both approaches should be vigorously pursued
so that this question may be settled. At the present time some
Schr\"odinger-based microscopic approaches utilize nucleon density
distributions
from Dirac-Hartree approaches which is inconsistent. On the other hand
the Dirac proponents should address open-shell target nuclei by some
approximation.
\item On the question of the influence of the bound-state problem upon
the medium-energy scattering problem this appears to be an only somewhat
tapped resource
at this time. Examples are the dispersion approaches and microscopic
approaches where in the former the single-particle levels are an important
constraint and in the latter the neutron and proton density distributions
define the folding volume. Given that some of the observables
used to select a ``correct'' bound-state approach are extremely
well measured (mass, rms charge radius, s.p. levels) means that the
bound-state problem can have tremendous influence
on the scattering problem. The converse is to some extent also true.
Thus, our recommendation is to vigorously pursue the influence of the
bound-state problem upon the medium-energy scattering problem
in mutually consistent approaches.
\item Some specific recommendations based primarily on the technical
presentations at this meeting are:
\begin{enumerate}
\item Work on the extension of the JLM folding model to higher energy
nucleon-nucleus
scattering should certainly be continued.
\item The various microscopic approaches should all include more detailed
comparisons of the calculated and measured integrated scattering observables,
proton total reaction cross sections and neutron total cross sections, as
functions of projectile energy and target nucleus, than in the past.
It is noted here that an accurate predictive capability for these observables
is extremely important for the medium-energy applied programs -- such as
the accelerator transmutation of radioactive waste.
\item Work on the extension of the dispersion approach to deformed nuclei
should certainly be continued.
\item The attractive medium-range $NN$ interaction is still not
understood and it needs a dedicated systematic attack. This is a topic
that is potentially rich in the physics payoff because it connects
the fictitious $\sigma$ meson, two pion exchange, the Walecka model,
QHD-I and QHD-II,
chiral symmetry, and QCD.
\item ECISVIEW should include a search package that provides the
running $\chi^{2}$ in a corner box of the interactive display.
[I understand that this has been accomplished just after the meeting.]
\item A global medium-energy nucleon-nucleus optical potential is,
and remains, an important goal in this field.
\end{enumerate}
\end{enumerate}
\noindent Although this meeting was advertised as addressing the
nucleon-nucleus optical model up to 200 MeV, the presentations and
discussions here addressed the model at energies as high as 1 GeV.
This is fortunate because some of the medium-energy applied programs
will require proton beams up to 2 GeV and, therefore, scattering
observables up to 2 GeV will have to be well understood. Thus,
we should begin addressing the extensions of experiment as well as
theory and modeling to the higher energies as soon as possible.
Note that a few GeV is where, at the present time, scattering formalisms
based upon partial wave expansions become intractable for
medium and large mass nuclei. For these reasons, it probably makes sense
for us to get
together again to assess where we are (1999 $\pm$ 1?).
In conclusion I thank the organizers for the privilege of summarizing
this meeting. \\
 
\noindent Los Alamos report LA-UR-96-4759
\end{document}